# Research and Development in Front-End Electronics for Future Linear Collider Detectors

Bruce A. Schumm

Santa Cruz Institute for Particle Physics and the University of California at Santa Cruz
schumm@scipp.ucsc.edu

## Abstract

To probe physics of interest a decade after the LHC begins to take data, the Linear Collider will need to have the capability of performing precision measurements of masses and couplings. We discuss the demands that this places on its associated detectors, and the status of and plans for continued R&D in front-end electronics that is necessary to meet these demands.

## I. INTRODUCTION

The physics case for the Linear Collider (LC) hinges on its ability to perform exacting studies of phenomena associated with electroweak symmetry breaking and the properties of the top quark. Running ten years after the first LHC data, the Linear Collider will make its impact through precision studies of new states (Higgs, supersymmetry) that are likely to be discovered at the LHC, and through measurement of gauge boson couplings that are precise and comprehensive enough to disentangle clues about states beyond the direct reach of either the LHC or LC.

Given the states that have been discovered so far, the rates of certain Standard Model processes, such as the scattering of W bosons, are predicted to diverge at energies above the electroweak scale. It is thus all but necessary that new physics – most likely that associated with electroweak symmetry breaking – will emerge as the LHC and LC probe the electroweak scale. Evidence for new states hinting at the physics of the grand unification scale, such as supersymmetric partners of Standard Model particles, may well also be captured within the LHC and LC data sets.

At the LC, the Standard Model (SM) Higgs can be studied in a model-independent fashion via the process $e^+e^- \rightarrow ZH$, with the subsequent decay $Z \rightarrow l^+l^-$ (see Fig. 1). Given the precise knowledge of the initial state, the Higgs can be detected as a clear peak in the invariant mass spectrum of the system recoiling against the Z, limited in resolution only by the intrinsic width of the Z. To fully exploit this channel, the tracker should be able to reconstruct the dilepton invariant mass with a resolution better than the Z width, a specification leading to a charged-particle transverse-momentum resolution $\delta p_\perp/p_\perp^2$ of $5 \times 10^{-5}$ or better, an order of magnitude more precise than that of any existing cylindrical geometry detector.

Should the Higgs be discovered at the Tevatron or LHC, it will be of utmost importance to determine if the Higgs is that of the SM, or that of a richer scenario such as the Minimal Supersymmetric Model. Thus, it will be of great interest to measure Higgs boson couplings, as manifested by its branching fractions to various quark and boson states, with great accuracy. The classification of the Higgs decay final states, particularly for the case of heavy quarks, requires the measurement of charged-particle impact parameter resolution to better than $3\mu m$ at asymptotically high momentum, degrading to no worse than $10\mu m$ at p = 1GeV/c.

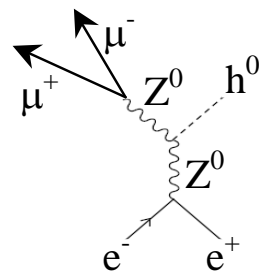

Figure 1: The Higgstrahlung diagram.

On the other hand, should the Higgs mechanism not be Nature's approach to electroweak symmetry breaking, our understanding will need to derive from the precise measurement of the *WW* scattering process (Fig. 2), for which it will be essential to distinguish between the *WW νν* and *ZZ νν* final states. To do this in a statistically acceptable fashion requires the inclusive separation of these two final states via the reconstruction of the combined invariant mass (`dijet mass') of the two jets emanating from the decay of each vector boson.

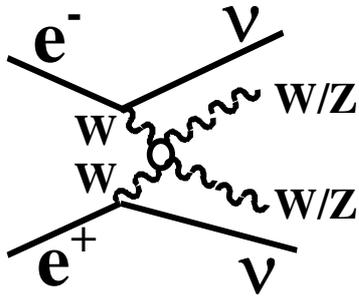

Figure 2: WW Scattering at the LC.

Fig. 3 shows a simulation of the dijet mass reconstruction for the $WW\nu\nu$ and $ZZ\nu\nu$ final states under two different assumptions for the jet energy resolution. It is seen that a jet energy resolution of $30\%/\sqrt{E}$ is necessary to consistently distinguish between these two final states.

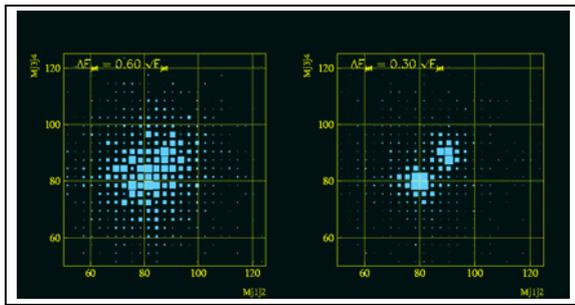

Figure 3: Calorimetric reconstruction of the $WW\nu\nu$ and $ZZ\nu\nu$ dijet mass under two different assumptions for the jet-energy resolution: $60\%/\sqrt{E}$ (left) and $30\%/\sqrt{E}$ (right) [1].

To achieve the goal of $30\%/\sqrt{E}$ for the jet energy resolution requires the implementation of the 'energy flow' approach to jet reconstruction, for which clusters from charged hadrons are removed from the calorimetric reconstruction, with the kinematic information coming instead from the much-more-precise tracking measurement. This approach places stringent demands on the Moliere radius and granularity of the calorimeter, with corresponding demands on the design of its readout electronics.

Very recently, the International Technology Review Panel has recommended the superconducting alternative for the acceleration of the LC beams. This approach features a collision repetition rate of 337 nsec, extending over a period of 950 msec (2820 pulses), and repeating with a frequency of 5 Hz. To avoid accumulating an intolerable occupancy in the detector components, it is desirable to provide timing that allows detector signals to be ascribed to a particular crossing, or a time window incorporating as few crossings as possible. Also, to avoid the need for active cooling, it is advantageous to exploit the intrinsic duty cycle of $5 \times 10^{-3}$ of the superconducting alternative. These are important considerations for the development of front-end electronics for the LC Detector, although not all efforts have yet had the time to incorporate this decision into their development work.

Finally, it should be mentioned that, surprisingly, time is somewhat pressing for the basic R&D phase of LC detector development. Assuming a ten-year period between the selection of detector technology options and the start of data taking, innovative detector R&D should ideally occur between ten and fifteen years before the LC turns on. In other words, the window of opportunity for developing detector and electronic components – components up to the task of exploiting the rich physics program of the LC – is now open.

## II. FRONT-END ELECTRONICS FOR LINEAR COLLIDER CALORIMETRY

Two major groups are active in the development of calorimetry for the LC. The CALICE collaboration is a sizable group comprised of 28 institutions from eight nations in Europe, North America and Asia. CALICE has a broad focus, including finely-granulated Silicon-Tungsten electromagnetic calorimetry, and the development of both gaseous- and scintillator-based readout for hadronic calorimetry. The Silicon Detector (SD) Silicon-Tungsten calorimetry group is composed of eight scientists and engineers from SLAC, Brookhaven National Laboratory, and the University of Oregon, and is concerned with the development of a finely-segmented Silicon-Tungsten (Si/W) calorimeter for an intermediate-sized detector that features all-silicon tracking.

In view of the demanding requirements of the energy-flow approach, the goal of Si/W development is to read out the electromagnetic calorimeter (ECAL) with full longitudinal segmentation, and with a transverse segmentation on the order of the ~1cm Moliere radius of tungsten. Such a calorimeter has never been built, and as a result our understanding of shower development on such a fine scale is not mature. Thus, a near-term goal of the CALICE group is to build and operate a ~1 m$^2$ prototype, to be comprised of 1 cm$^2$ planar silicon diodes interleaved between tungsten sheets of thickness between 1.4 and 4.2 mm, featuring approximately 10,000 channels of readout.

To meet the needs of the CALICE ECAL prototype, the LAL Orsay electronics design group has developed the FLC_PHY3 ASIC [2]. Nominally incorporating 13 bits of linear dynamic range, the FLC_PHY3 has 16 preamplifier gain settings, ranging over a factor of 15, to accommodate variations in detector and preamplifier response, that can be selected prior to operation. After preamplification, the signal is split into separate x1 and x10 shaping stages that impose a 200 ns peaking time, and then presented to individual track-and-hold stages that can be selectively read out by the data acquisition system. The Equivalent Noise (ENC) for the high-gain (small-signal) channel has been measured to be $ENC = 950 + 34 * C_{load}(pF)$ electrons.

While the CALICE ECAL prototype is expected to help assess the feasibility of the energy-flow approach, and substantially refine our understanding of shower development, many of the engineering and integration issues that will confront the designers of a full-scale system will remain un-addressed. To preserve the Moliere radius, it is necessary to avoid large low-Z gaps between tungsten layers, which implies that the front-end electronics (including zero-suppression, multiplexing, addressing and time stamping) will need to be integrated into the active region of the detector. In addition, the power dissipation will need to be minimized, in order to avoid the material and complexity associated with active cooling. In addition, a full-scale Si/W ECAL for the LC Detector will contain between 10 and 100 million channels.

Both the CALICE and SD calorimetry groups are addressing these issues. Fig. 4 shows the tentative solution proposed by the SD group. The 1mm gaps between tungsten plates are tessellated by hexagonal diode detectors with a lateral dimension of 5mm. Traces from 1,024 hexagonal pixels present sensor signals to a compact bump bond array that is connected to a single, 1,024-channel readout ASIC. Service and IO lines are carried on multi-layer PC boards that fill the remainder of the 1mm gap. A plug of material with high heat conductivity provides a thermal path from the readout ASIC to the tungsten plate just above; the tungsten then acts in bulk to carry the heat out of the body of the calorimeter, and is cooled only at the extremities of the calorimeter.

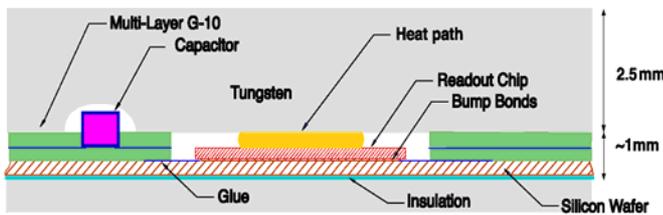

Figure 4: The readout system proposed for the SD ECAL.

The SD Si/W readout ASIC, currently under development, will feature rapid power-cycling so that its microcircuitry is energized only during the $5 \times 10^{-3}$ duty cycle of the LC, which is essential if the heat load is to be kept small enough that the calorimeter can be cooled only at its extremities. The ASIC is also being designed to cover a dynamic range of up to 2,500 minimum-ionizing depositions per pixel, and employs a x20 'dynamical' gain switch that is implemented automatically within the microcircuitry when the pulse-height crosses a preset threshold. This ASIC is expected to be fabricated in late 2004; a similar effort, on a similar timescale, is underway within the CALICE group.

Although the energy-flow approach de-emphasizes the need for good energy resolution for hadronic calorimetry, the requirement of identifying and removing clusters from charged hadrons retains the need for high granularity in the hadronic calorimeter. CALICE is pursuing two directions in its development of hadronic calorimetry for the LC Detector: a steel/scintillator sandwich with either 5x5 cm$^2$ ('tile HCAL') or 3x3 cm$^2$ ('semi-digital HCAL') tiles, and steel radiator with gaseous readout with a pixel size of 1x1 cm$^2$ ('digital hcal').

A recent development in the readout of visible-light signals (such as those produced by wavelength shifting fibers embedded in the scintillating tiles) is the invention of the silicon photomultiplier (SiPM) [3], shown in Fig. 5. Each 20x20μm$^2$ pixel is run in Geiger mode, with a gain of ~10$^6$. Each SiPM chip contains of order 10$^3$ Geiger pixels and mates to a single wavelength shifter fiber. Thus, intensity information is derived from the number of pixels that experience Geiger breakdown, rather than the size of the signal from individual pixels. Signal from individual pixels are ganged together on the SiPM chip and read out by a single electronics channel.

The unique calibration opportunity provided by a SiPM-based readout is being incorporated in the design of a front-end ASIC under development at DESY. The single- and few-photoelectron response provides a powerful absolute calibration scale to track the gain and (combined with accumulated minimum-ionizing signals, which average approximately 25 photoelectrons) small-signal linearity of the SiPM readout. However, to avoid confusion from overlapping single photoelectron signals, the front-end electronics must employ a short shaping-time readout, with relatively high gain. For the data-taking mode, on the other hand, a long shaping-time response will be employed to minimize noise and to integrate the full signal duration from the scintillator signal. The DESY group has proposed two possible approaches for the calibration mode: an externally selectable gain and shaping time, or a setting that eliminates the undershoot of the front-end response ('unipolar response'), thereby greatly shortening the recovery time of the circuit. The design of a prototype ASIC incorporating both alternatives is nearing completion, and should be submitted shortly.

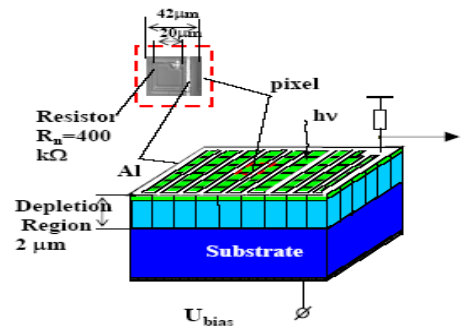

Figure 5: The Silicon Photomultiplier (SiPM) is a ganged set of Geiger-mode pixels implemented on a single silicon substrate.

The detector options that have been proposed for the gaseous readout option (RPC, Geiger or streamer tubes) are all fairly well understood; the primary challenge is the large channel count implied by the 1x1cm$^2$ segmentation. A group based at Fermi and Argonne National Laboratories is designing an ASIC [4] to readout RPC detectors for a CALICE digital HCAL prototype. This ASIC gangs together 64 channels, reporting out the time and pattern of hit pixels in a four-word code at a rate of 10 MHz. It can be programmed to suppress individual noisy channels, and has an optional 2 μs pipeline for accommodating triggering. This prototype ASIC is expected to be available for testing in early 2005.

## III. FRONT-END ELECTRONICS FOR LINEAR COLLIDER TRACKING

There are many advantages to using a large-volume TPC for the central tracking device for the LC Detector. Taking into account the z-coordinate information provided by the drift, the TPC is a fully three-dimensional tracker. Although the ~100 μsec drift time will provide an intrinsic integration over several hundred beam crossings, a comparison of reconstructed tracks with segments from the vertex detector will allow tracks to be assigned to a unique beam crossing, limiting confusion from overlapping events. The high (~4T) field necessary to keep electron/positron pairs from the beam-beam interaction from entering the vertex detector would also act to limit transverse diffusion in a TPC, making this approach to central tracking inherently precise.

The wire-chamber readout scheme that has instrumented past and existing TPC's will not provide a point-resolution precise enough to exploit the intrinsic accuracy of a TPC operated in a high magnetic field. Instead, TPC development groups are exploring the use of micropattern gas detectors, such as GEM's and Micromegas, that produce compact, localized amplification that is sensed by conducting pads with a typical readout pitch of 1-3 millimeters.

A number of groups have developed prototype TPC facilities, and are beginning the process of optimising the micropattern detector properties for use in the readout of a large-scale TPC at the LC. To this point, these groups have made use of electronics developed by the STAR collaboration [5] consisting of a sixteen-channel, two-chip sequence that provides preamplification, shaping, an analog pipeline and analog-to-digital conversion.

There are several attributes of a TPC optimized for use at the LC that will require the further development of front-end electronics. To suppress ion feedback into the TPC tracking volume, it will likely be necessary to run the micropattern gas detectors with relatively low gain, thus requiring attention to amplifier noise. To exploit the resolution limit in the z-coordinate, the charge digitization will need to be performed by flash ADC's with a 100 MHz sampling rate. The reduction in the readout pitch relative to previous (wire-chamber) readout schemes will increase the number of channels by an order of magnitude. On the other hand, the endplate regions of existing TPC's incorporate a large fraction of a radiation length of readout and support material, which would substantially degrade the quality of information entering the precision low-angle calorimetry of the LC Detector.

Thus, substantial attention will need to be paid to limiting the material associated with the end-plane readout scheme, in the face of a vastly increased data throughput. Substantial effort will need to be put into areas such as zero-suppression, buffering, waveform processing, and power cycling as the TPC front-end electronics is designed. This effort should be getting under way shortly as the various TPC prototyping efforts begin to pin down the layout and operating parameters of the micropattern gas detectors.

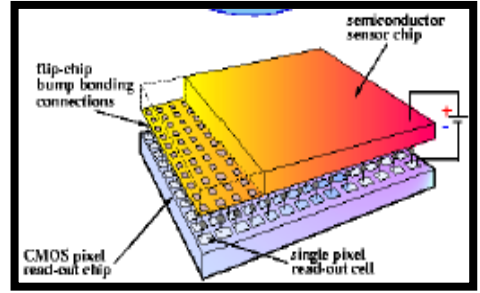

Figure 6: The Medipix2 CMOS pixel sensor [6].

An independent approach to the electronic readout of micropattern gas detectors has been proposed by the LC tracking groups at NIKHEF and Saclay. Fig. 6 shows the 55x55 μm$^2$ Medipix2 CMOS pixel sensor [6], which has been configured by the NIKHEF and Saclay groups to sense individual avalanches from GEM and Micromegas detectors reading out a small prototype TPC. Fig. 7 shows a segment of a cosmic ray, with an associated delta ray, that was read out with the Medipix scheme. Avalanches induced by individual ionization clusters are visible, implying that this readout scheme provides the most possible information about the passage of the track through the detector volume. Such a readout scheme would promise to contribute to optimal space-point and dE/dX resolution. On the other hand, instrumenting the TPC with such a system would imply roughly ten billion readout channels, five orders of magnitude more than currently operating chambers.

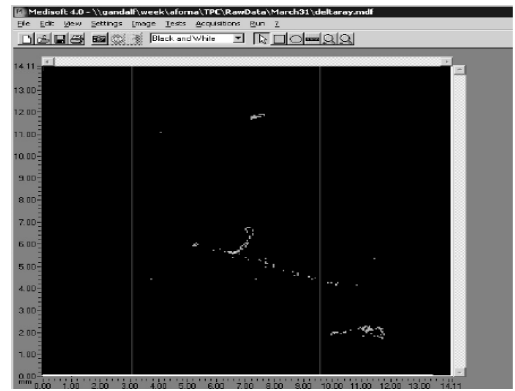

Figure 7: Readout of a micropattern-instrumented TPC prototype with the Medipix2 pixel sensor.

An alternative to gaseous tracking for the LC Detector is a central tracker composed of five to ten layers of silicon strip detectors. Typical solid-state tracking detectors employ 1.5–2% of a radiation length of material per layer for sensors, electronics, mechanical support, power and data traces, and cooling lines. For the LC Detector, for which precision is essential, this amount of material would significantly compromise the momentum resolution, and degrade the information entering the calorimeter.

Groups at LPNHE Paris and the University of California at Santa Cruz are exploring the possibility of a long shaping-time readout that would allow the instrumentation of detector segments as long as 167 cm (the half-length of the outermost layer in the SD detector design) as a single ladder. Such a detector could be read out at its ends only, greatly reducing the need for servicing in the tracking volume. In addition, the implementation of power cycling (turning the electronics off in the 200 msec gap between pulse trains) could be used to reduce the total power dissipation of such a solid-state tracker to approximately 10 W, eliminating the need for active cooling, and avoiding the material that would otherwise be associated with a cooling system.

Fig. 8 shows the curvature resolution ($\delta p_t/p_t^2$) for a five layer solid-state tracker (SD Thin) whose only material is that of the sensors themselves, which are assumed to be 300 μm thick in the outer layers, tapering to 100 μm in the innermost layer. While not a realistic design, this exercise shows that there is substantial room for improvement at low and intermediate momentum relative to a more conventional solid-state tracker (SD Thick). In addition, it may well be possible to design a solid-state tracker that is competitive in curvature resolution with gaseous tracking alternatives (L) over the full range of track momentum. Simulations done by the Santa Cruz group suggest that, for 3 μsec shaping time, an operating point for a 167 cm long ladder can be developed that achieves 99.9% hit efficiency while suffering a noise occupancy of only 0.1%.

Roughly speaking, the temporal resolution of such a device is given by

$$\sigma_t \approx \frac{\tau_{shape}}{SNR}$$

where $\tau_{shape}$ is the characteristic shaping time, and SNR the overall signal-to-noise ratio of the preamplification. For a shaping time of 3μsec and a signal-to-noise of 12, this corresponds to a temporal resolution of approximately 250 nsec, leading to the possibility of beam-pulse-by-beam-pulse discrimination of individual silicon clusters.

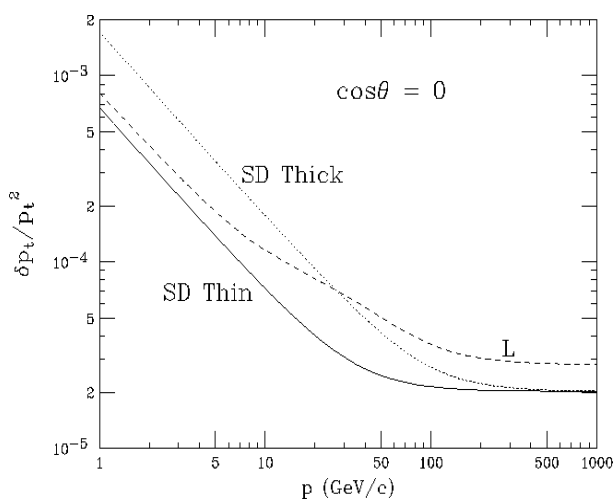

Figure 8: Curvature resolution of an idealized low-mass solid-state tracker (SD Thin), relative to conventional solid state (SD Thick) and gaseous (L) tracking alternatives.

The LPNHE and Santa Cruz groups are independently designing power-cycled, long shaping-time ASIC's to explore the feasibility of the thin all-silicon-tracker approach. Both groups expect to have chips to study by the beginning of 2005. Electronic simulations done by the UC Santa Cruz group indicate that it is possible to stabilize a circuit for operation within 100-150 μsec of switching it on, suggesting that the full 99.5% duty-cycle power savings may be achievable. The Santa Cruz pulse-development simulation also suggests that, with the use of a lower secondary threshold, read out for nearest-neighbors only when the higher primary threshold is exceeded, a point-resolution of better than 7 μm can be achieved, as assumed for the resolution study of Fig. 8.

In addition to the limited material in the central region, such a tracker, read out only at its ends and communicating with the outside world via high-bandwidth optical fibers, would also place relatively little material in the way of the forward calorimetry – an essential goal if precision calorimetry is to be done outside the central region.

## IV. FRONT-END ELECTRONICS FOR LINEAR COLLIDER VERTEXING

The international baseline vertex detector concept is shown in Fig. 9. The baseline technology is 22x22 μm$^2$ CCD pixels, with a space-point resolution of 5 μm, and with an overall material burden of 0.12% $X_0$ per layer.

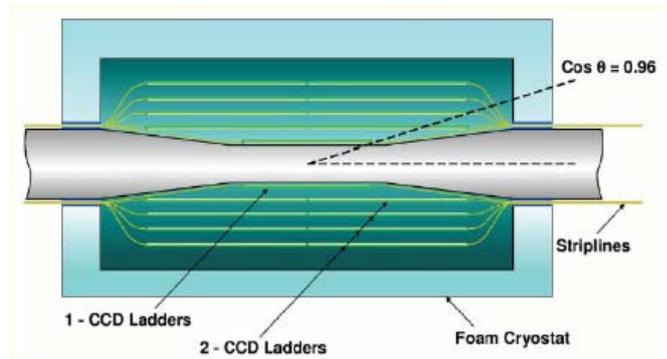

Figure 9: The international baseline design of the Linear Collider Detector Vertex Tracker.

Immersed in a magnetic field of 4-5 T, which forces most of the $e^+e^-$ pairs from the beam-beam interaction into a very tight cone about the beam axis, the inner layer of the Vertex Tracker is expected to reside 1.2 cm from the beam trajectory. Altogether, the impact-parameter resolution expected for such a device (Fig. 10) is between a factor of 2-3 better than that of the SLD Vertex Detector, the most accurate existing cylindrical-geometry vertex detector.

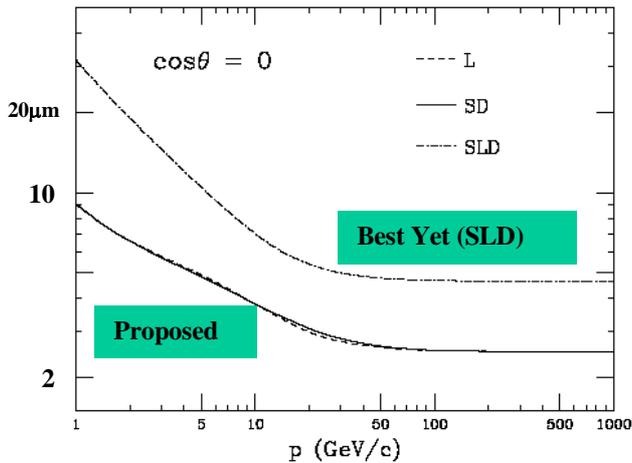

Figure 10: Impact parameter resolution (in μm) versus momentum for the best existing vertex detector (SLD) and the proposed Linear Collider Vertex Tracker.

The application of CCD technology to the high-energy LC is not immediate. The SLD ran with a single bunch-crossing every 8.5 msec, while the LC will employ almost 3,000 bunch crossings per train. With a single readout pad per CCD sensor, the detector would integrate over the full train, leading to intolerable backgrounds.

To this end, the Rutherford group, along with the international Linear Collider Flavor ID (LCFI) group, has worked with Britain's E2V Technologies to develop a 'column parallel' CCD that is read out through numerous pads – one for each column of the CCD. The prototype sensor has been manufactured with three different implementations of the readout: straight, unbuffered readout column designed to mate to a charge-sensitive amplifier, and readout columns buffered by either one or two on-board transistors (the latter being equivalent to the readout structure of the traditional CCD) designed to mate to a voltage-sensitive amplifier.

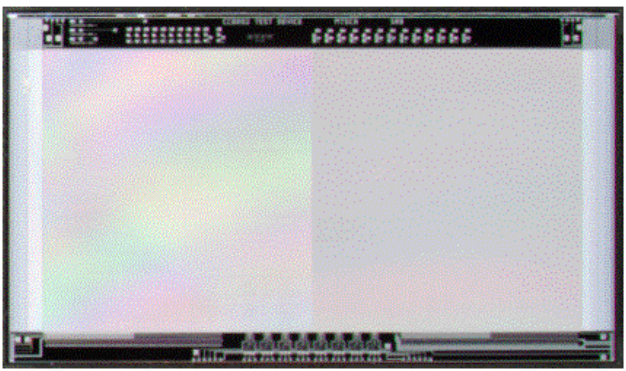

Figure 11. The LCFI/E2V CPC-1 column-parallel-readout CCD prototype.

The prototype column-parallel readout is shown in Fig. 11, while Fig. 12 shows the Column-Parallel Readout 1 (CPR1) ASIC designed by the group to read out the charge-sensitive and single-buffered voltage-sensitive pads of the CPC-1 sensor (two regions of charge- and voltage-sensitive outputs can be seen on either side of a central region of double-buffered voltage-sensitive readout pads). Minimum-ionizing signals have been observed in the single-buffered voltage sensitive readout, and the group is now in the process of assessing signal-to-noise.

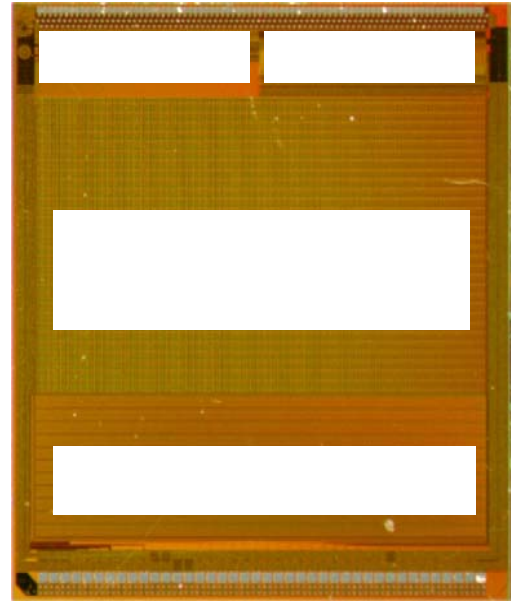

Fig 12. The CPR1 column-parallel CCD readout prototype ASIC.

Lately, the LCFI group, based on experience with RF pickup in the SLD VXD tracker, has become concerned about the feasibility of reading out a CCD detector during the beam spill. Since it is imperative to do so (to avoid intractable occupancy), the group has begun to explore the feasibility of using Image Sensors with In-Situ charge storage (ISIS) to avoid having to cope with RF noise during the sensor readout.

The ISIS principle is diagrammed in Fig. 13. Charge accumulated from the epilayer during a set time period is transferred into an n-channel storage well near the surface of the pixel. After another period of the same duration, this charge is shifted one storage well to the right, while the charge accumulated during this second period is shifted into the first storage well. In this way, the charge is accumulated within the pixel, without attempting to read it out, for the full duration of the pulse train. However, the train is divided up into ten or so temporal buckets according to which storage well it finds itself in as it is finally read out after the noise from the beam crossings has subsided.

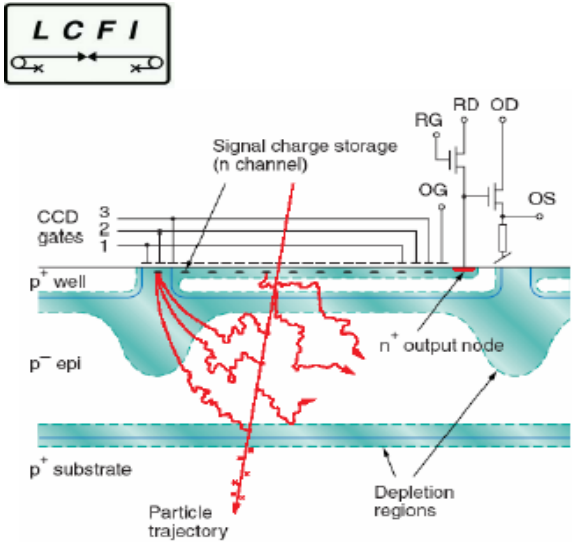

Figure 13: Diagram of the ISIS in-situ charge storage scheme. Shown is a single pixel; the n-channel storage wells act to subdivide the charge accumulated during the full beam spill into distinct temporal slices.

Exploration of ISIS sensors for application to the LC Vertex Tracker is in its early stages. Hope that this technology can be developed and refined in time for such an application hinges in part on its industrial applications.

By effectively proposing to place a shift register on the surface of the ISIS sensor, the boundary between the arenas of detector R&D and electronic readout R&D begins to become obscured. This trend is even more evident in the last class of devices that will be discussed in this paper: Monolithic Active Pixel Sensors (MAPS).

Hybrid active pixel sensors, read out by dedicated ASIC's that are bump-bonded to the pixel sensors, are now enjoying widespread application, particularly in the high-radiation environment of the LHC detectors. For a number of reasons, though, these hybrid detectors are not likely to achieve the resolution of the LC Detector baseline design, shown in Fig. 10. Pixel sizes are typically of the order of 100x100 $\mu m^2$ (compared to 22x22 $\mu m^2$ for the SLD Vertex Detector), limiting the point resolution that can be achieved. The fill-factor of the independent readout ASIC is approximately one, and it is bump bonded to the pixel sensor, leading to a substantial material burden. Finally, the hybrid detectors require active cooling, adding further material, and significantly limiting the resolution in the intermediate-momentum range that is critical for precise flavor tagging.

As a result, a number of laboratory and university groups have begun to explore the idea of Monolithic Active Pixel Sensors (MAPS), for which much of the readout circuitry is deposited directly onto the pixel sensor itself, within the epitaxial and metal layers. By avoiding bump bonds, the pixel dimensions can be kept small – on the order of the 22x22 $\mu m^2$ of the CCD sensors. With thinning of the substrate, column-parallel readout addressing, and power cycling, it is conceivable that one or more MAPS schemes will be able to approach the exquisite resolution depicted in Fig. 10.

A group with the LEPSI electronics collaboration at IRES Strasbourg has produced a sequence of monolithic pixel sensors collectively known as the MIMOSA series. Early MIMOSA sensors sprang from commercial development of high-end digital photography, and employed minimal on-pixel circuitry: the MIMOSA V, which features a 17x17 $\mu m^2$ pixel dimension, incorporates a two-transistor amplification stage and a single CMOS row-select switch, and was fabricated in 0.6$\mu m$ AMS CMOS. The MIMOSA V is likely to be employed as a baseline for the STAR experiment vertexing upgrade. For LC applications, though, substantially more electronic functionality will need to be incorporated into individual pixels, including correlated double sampling, power cycling, thinning of the sensor bulk, discrimination and zero suppression, and column-parallel readout architecture. Nonetheless, with the deep-submicron processes currently available, it is hoped that this functionality can be achieved while maintaining a pixel dimension on the order of 25x25 $\mu m^2$.

A second interesting monolithic approach is the DEPFET concept, under development by collaborating groups at the Universities of Bonn and Mannheim, and shown schematically in Fig. 14. Charged released by the passage of ionizing radiation is collected in an n-well at the gate of a CMOS transistor that has been deposited on the surface of the pixel. This charge acts to alter the gate voltage, steering current through the transistor proportional to the modified voltage, and thus to the amount of accumulated charge. While such devices are now beginning to sense the passage of ionizing particles, separate preamplifier/processing and addressing ASIC's must be employed in order to read out and provide the functionality necessary for application to a LC Detector. Substantial work remains before such an approach could be employed in a low mass, hermetic, and robust LC Vertex Tracker.

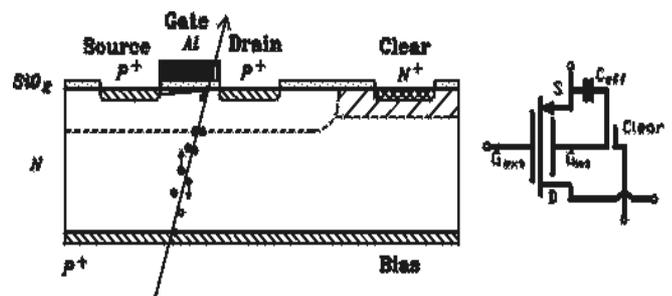

Figure 14: Schematic design of the DEPFET monolithic active pixel sensor; the effective electronic circuit is shown to the right.

Another approach is that of the Silicon on Insulator (SOI) technology, shown schematically in Fig. 15. A high-resistivity n-type substrate is depleted by a $p^+$ impregnation, leading to charge collection through the majority of the bulk, and subsequently large signals. As for other monolithic approaches, readout circuitry is grown directly onto the sensor epitaxial layer, with the advantage that both senses (PMOS and NMOS) of CMOS structures are now possible. Again, a substantial period of development lies between current implementations and application to the LC Detector.

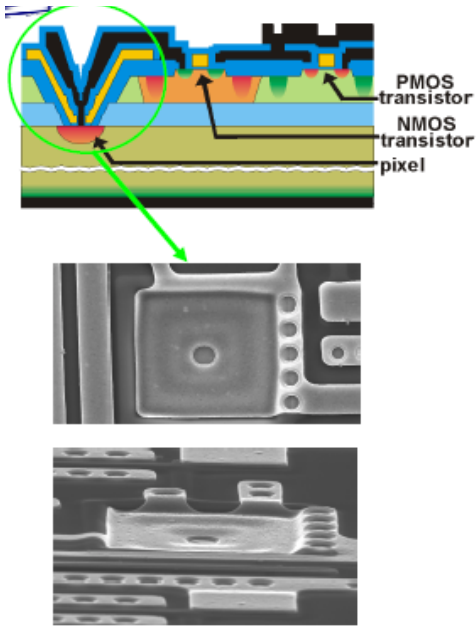

Figure 15: Schematic of the Silicon-on-Insulator (SOI) approach.

## V. SUMMARY

To exploit the full potential of the Linear Collider, it is necessary to design and construct a detector with performance characteristics that exceed those of other existing and proposed cylindrical geometry detectors. The development of precise Linear Collider detector technologies must happen in concert with – indeed, in some cases, must be enabled by – the development of advanced readout circuitry. In most cases, these developments are in directions quite dissimilar from R&D activity geared to hadron colliders. Furthermore, the window of opportunity for this basic R&D period for the Linear Collider is now upon us; commensurately, a large number of electronics development efforts are underway in all regions of the particle physics community.